# Soft X-ray microflares

**I. K. Mirzoeva**

Space Research Institute, Russian Academy of Sciences, ul. Profsoyuznaya 84/32, Moscow, 117997, Russia
e-mail: colombo2006@mail.ru

**Abstract**

Soft X-ray solar bursts are studied within the framework of the Interball–Tail Probe project with the RF–15I–2 solar X-ray photometer–imager. The low-intensity microflares observed in September–December, 1995 are analyzed. Weak bursts with powers up to $10^{-8}$ W/m$^2$ were detected. All the events were confirmed by GOES observations. Parameters of these microflares are determined. A physical mechanism for the low-intensity solar events is discussed.

The intensity distributions of the microflares are compiled. The energy distribution of solar flares is concluded to deviate from a power law. The lower limit of the solar-flare fluence distribution is found. This conclusion is verified based on RHESSI data.

Correlations between the diurnal mean peaks of X-ray bursts of different classes and the daily means of the thermal background emission from the solar corona are revealed.

**1. Introduction**

The vast majority of works devoted to solar flares are dedicated to strong, high-flux solar events. This is explained mainly by the capabilities of instruments used for observations. Recent progres in instrumentation opens up new possibilities for studying small-scale, low-flux solar events.

Note that the low-intensity solar events are conventionally classified according to their total energy release not exceeding $10^{27}$ erg. In addition, class-A bursts according to the GOES classification, whose maximum X-ray powers are less than $10^{-7}$ W/m$^2$, can also be considered low-intensity solar events.

Since 1969 till nowadays the solar flares are classified according to their (peak) X-ray fluxes at the Earth orbit. The X-ray flux of a flare is measured in watts per square meter (W/m$^2$). The generally adopted classification of solar flares in the X-ray range 2–15 keV is summarized in Table 1 below.

Table 1

| Flare class | Burst flux (W m$^{-2}$) |
|---|---|
| X | $10^{-4}$–$10^{-3}$ |
| M | $10^{-5}$–$10^{-4}$ |
| C | $10^{-6}$–$10^{-5}$ |
| B | $10^{-7}$–$10^{-6}$ |
| A | $10^{-8}$–$10^{-7}$ |

This scale has been adopted in the GOES project in the mid-1980s and is successfully used at present [1].

According to Table 1, the lowest-intensity solar events are class-A flares with peak X-ray fluxes of $10^{-8}$–$10^{-7}$ W/m$^2$. Such solar flares are mostly observed in soft X-rays, i.e., in the photon energy range 2–15 keV.

The low-energy X-ray flares in the photon energy range 3–12 keV were analyzed in [2] with the RHESSI spectrometer designed for detecting low-energy solar X-rays and imaging the solar disk. It was shown in that work that each of 7 X-ray microflares observed within one hour is a superposition of thermal and non-thermal components and has a fairly low energy release of $10^{26}$–$10^{27}$ ergs. The most important works on the microflares sphere during the last several years are [3], [4].

Parameters of the so-called nanoflares were studied in [5]. For example, some authors treat ultraviolet enhancements at wavelengths 171 and 195 angstroms observed even in quiet-Sun years as one of the weakest flare-activity manifestations. A classification of virtually all solar events in the wide energy range $10^{24}$–$10^{32}$ erg was also proposed in [5]. Figure 1 shows the composite diagram illustrating the classification of solar events with various fluxes. This diagram was plotted using the results by different authors. Here, the total energy release and the flare rate are reckoned along the $x$ and $y$ axes, respectively. It is believed that any flare event can be classified into one of the following classes: milliflares (or simply flares), microflares, and nanoflares. In accordance with [5], a flare is characterized by an energy release of $10^{29}$–$10^{32}$ ergs. It is responsible for both hard and soft components of the X-ray emission. A microflare energy release is $10^{27}$–$10^{29}$ erg. The energy release of a nanoflare is below $10^{26}$ erg. These events can be observed only in the ultraviolet. It is pointed out in that work that there is a "gap" in the energy range $10^{26}$–$10^{27}$ erg caused by the absence of experimental data.

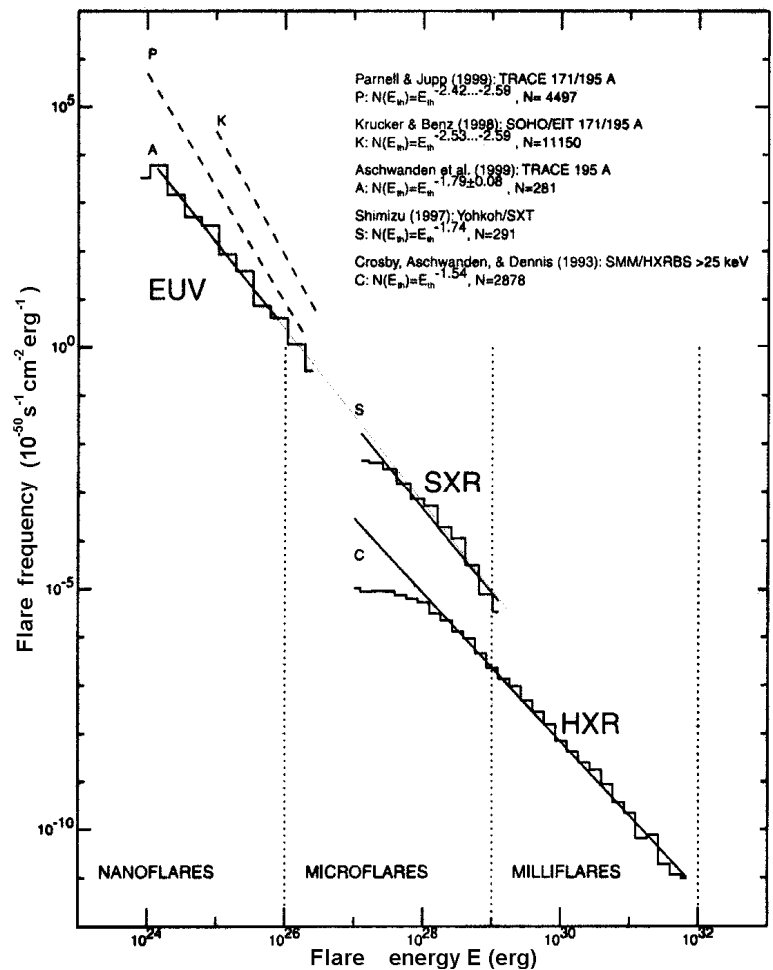

Fig. 1. The flare rate as a function of energy.

The existence of even weaker events called picoflares in the range of extremely small energy releases of $10^{17}$–$10^{24}$ erg is also discussed in the same work. According to authors' viewpoint, any local sporadic energy release on the Sun is a flare. In this case, even simple heating of the solar plasma due to, e.g., the formation or evolution of extremely small magnetic structures in the chromosphere, which typically results in an enhancement of ultraviolet emission, should be called a flare. However, such an approach is not undoubted.

In general, the problem needs further revision and is currently conventional to certain extent. On the one hand, it is clear that the energy release *per se* can be continuus, so that it is yet difficult to point out a factor constraining the possible energy release from below. On the other hand, it seems expedient to border the energy release from below thus distinguishing flares against continuous energy release.

In our viewpoint, a "flare" is such a local solar energy release that results in forming some threshold number of hot or accelerated electrons and thereby gives rise to X-ray emission (a burst) of thermal or bremsstrahlung origin. Such particles can be accelerated in local electrical fields which, in turn, can be generated along with the development of instabilities and current-sheet tearing in active regions. Therefore, a flare is the result of such a restructuring of plasma-magnetic configuration of an active region that causes the above-described processes. This concept will be accepted thoughout the present work.

Moreover, a flare is an integrated process related to a plethora of coexisting physical phenomena. That is why it is so difficult to reveal cause-effect links in the flare mechanism. In this respect, analyzing low-intensity flares is especially important since it provides for an opportunity to minimize integral effects and outline more definitely the sequence (chain) of physical phenomena in the flare development.

One of the most important results of this work is answering the question on the existence of the weakest solar events featuring peak X-ray fluxes below $10^{-8}$ W/m$^2$ [6] i.e., the existence of a class of weak X-ray bursts below the level indicated in Table 1 showing the classification of X-ray bursts. This class is referred to as class 0 hereafter.
Recently, these low-intensity flares have also been observed by other investigators, according to the RHESSI data [4], fig 7.

The lower limit is of interest for X-ray burst observations since according to present views, the explosive phase of a flare can be a sequence of elementary flare bursts [7], [8].

For example, in [7], a flare is considered a superposition of a few elementary energy-release events.

Thermal X-ray background of the solar corona and its relationship with low-intensity solar events are poorly studied at present. Advances in this line of research can shed light to the problem of solar corona heating.

It was assumed in work by [9] and [10], that the correlation between the solar X-ray background intensity and an X-ray microflare generated by particles accelerated during its onset in an active region stems from the absence of distinct separation of these two solar-emission components. Consequently, it was assumed that microflares and their host active regions (plasma-magnetic configurations) are linked much tighter with coronal plasma-magnetic configurations than active regions responsible for prominent flares. A number of researches drew attention to this fact. In particular, this issue was addressed for the first time in work by [11].

No relation was found between prominent class-C, -M, and -X flares and fluctuations of the thermal solar-corona background. It was also assumed in [2] and [7] that prominent flares represent secondary formations constituted by the superposition of a number of elementary energy-release events and generate all the following integral effects such as particle acceleration resulting in hard and soft X-ray, UV, and radio emission, as well as magnetohydrodynamic shocks.

## 2. Solar X-Ray Photometer-Imager RF-15I-2 Aboard The Interball-Taill Probe

INTERBALL is the solar-terrestrial programme aimed to study various plasma processes in the Earth magnetosphere by the system consisting of two pairs (satellite-subsatellite) of spacecraft above the polar aurora and in the magnetospheric tail respectively. Each pair of spacecraft consisted of a [Prognoz](#)-type satellite and smaller [Magion](#)-type subsatellite. The INTERBALL was a part of of the Programme coordinated by the [Inter-Agency Consultative Group (IACG)](#) for Space Science consisting of representatives of the European Space Agency (ESA), the National Aeronautics and Space Administration (NASA), Russian Space Agency (RKA) and Japan Institute of Space and Aeronautics Sciences (ISAS). According to this Programme a system of ten core spacecraft of the above agencies was to be spatially distributed between the L1 and L2 Sun-Earth libration points to study solar-terrestrial relationship.

The INTERBALL Project was designed to study various plasma processes in the circumterrestrial space as the principal way to study solar-terrestrial physical processes. The Project consists of two pairs (satellite-subsatellite) at high altitude orbits: to 200 000 km for the Tail Probe pair and to 20 000 km for the Auroral Probe pair. The inclination for both is 62.8 degrees. The Tail Probe is launched to an orbit with a low angle with the ecliptic plane to reach high altitude cusp and subsolar magnetopause regions on the dayside and then, the neutral sheet in the nightside tail. The Auroral Probe orbit is optimised for the magnetic conjunctions between the two pairs of satellites around the midnight meridian. This will allow to study the cause-and-effect relationships between the plasma processes in the tail and in the auroral particles acceleration region above the auroral oval with a high time-space resolution. In the other phase of the Project various solar wind disturbances and X-ray emission bursts on the Sun will be studied from the Tail Probe together with their effects on the cusp-cleft and other regions of the dayside magnetosphere and ionosphere measured simultaneously by the Auroral Probe. This again will allow to observe and analyse the cause-and-effect relationships in the solar wind/magnetosphere interactions [12].

Within the frames of the INTERBALL project there were carried out different researches concerning soft and hard X-rays (2-240 keV) from solar flares.

The primary purpose of the RF-15I-2 instrument is to measure the hard and soft X-ray components of solar-flare emission. This photometer–imager includes two units: the detector unit and the onboard-computer unit. Two sections can be identified in the detector unit: the hard X-ray photometer collecting photons from the entire solar disk with high time resolution and the soft X-ray photometer.

The photometer section of the detector unit (see Fig. 2) comprises detectors of three types: a soft X-ray proportional counter (PC) marked D in Fig.2, a hard X-ray scintillator detector (SCD, B in Fig. 2), and three semiconductor charged-particle detectors (A1, A2, and A3 in Fig. 2). The proportional counter is filled with a mixture of 90% Ar + 10%$CO_2$ under a pressure of 350 Torr and consists of two equal parts enclosed in the common volume. One part of the proportional counter has a 100-micron-thickness beryllium window with an area of 0.05 cm$^2$. The window of the second part is covered with $Fe^{55}$ which makes it possible to calibrate this detector permanently during the experiment. The scintillator detector consists of 8-mm-thickness and 50-mm-diameter NaI(Tl) crystal and a photomultiplier. The background is rejected with a special passive multilayer massive shield. A metal plate covered with the radioactive $Am^{241}$ can be moved out of the multilayer shield and used to calibrate the scintillator in the course of the instrument operation.

Eight energy channels of the proportional and scintillator detectors are used for amplitude analyses.

| Channel | Energy range, keV |
| --- | --- |
| PC1 | 2–3 |
| PC2 | 3–5 |
| PC3 | 5–8 |
| SCD1 | 10–15 |
| SCD2 | 15–30 |
| SCD3 | 30–60 |
| SCD4 | 60–120 |
| SCD5 | 120–240 |

The instrument is controlled by the PRAM onboard computer. In particular, the detectors and electronics are tested and the instrument is calibrated by a built-in radioactive source in the course of the start sequence initiated upon the switch-on. The data acquisition during in-orbit operation is PRAM-controlled according to their feasibility. The time resolution is software-controlled according to the current count rate. The high-voltage circuits of the photomultiplier and proportional detectors were switched off during the radiation-belt passages. The time resolution of the hard X-ray channels (SCD1–SCD5) can be tuned to 0.1 s and is fixed at a level of 2 s for the PC1–PC3 channels. The data by three semiconductor detectors are used to prompt the entrance of the spacecraft into the radiation belt. In addition, these data can also be used for qualitative evaluation of high-energy particle fluxes accompanying strong flares.

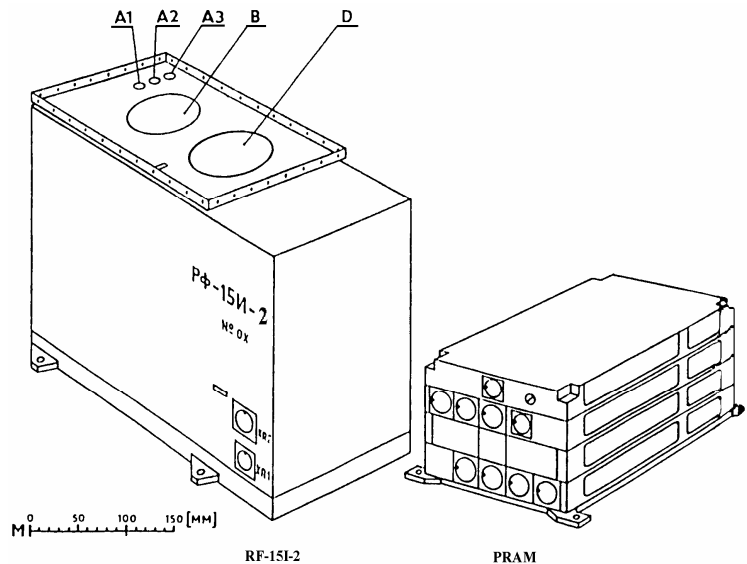

Fig.2. Solar X-ray photometer–imager RF-15I-2.

**3. Characteristics Of Weak Solar Flares In The Soft X-Rays**

In total, about 18,000 solar X-ray flares with various energy releases have been detected with the RF-15I-2 instrument in 1995–1999. This work is focused on a detailed analysis of the low-intensity solar events accordingly selected from the large field of the experimental data volume.

A detailed analysis of the selected data on X-ray bursts observed in 1995–1999 revealed a few parameters of extremely-low-flux events [6].

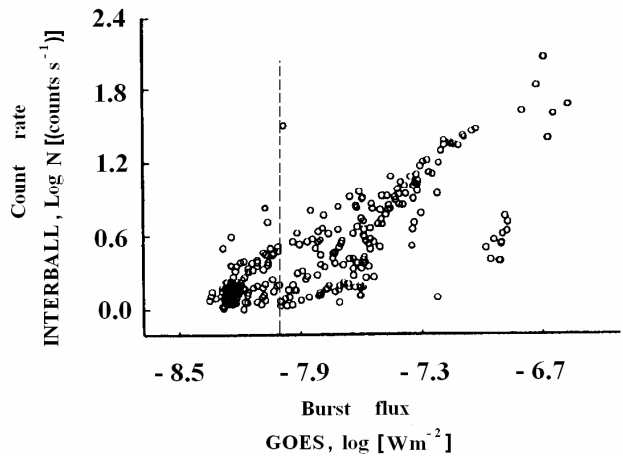

Fig 3. Comparative diagram of weak solar X-ray bursts in the photon energy range 2–15 keV based on the events observed by both GOES and INTERBALL spacecrafts in September–December 1995.

We have studied the following: total quantity of "class 0" microflares within the period of observation - September-December 1995, a maximum value of the X-ray flux for each "class 0" solar microflare in $W/m^2$, a total duration of each microflare, a duration of each microflare onset and decay phase, a daily mean background level in each event.

Characteristics of weak solar X-ray flares are studied for the most favourable observation period covering 9th to the 12th months of 1995 when a transition from the solar maximum to the solar minimum began and the number of flares in the selected photon energy range was not very large. This minimized the number of overlapping events. Then the periods of radiation-belt passages by the satellite were excluded from the data. As a result, the following dates free of high-flux solar events were selected:

September 2, 3, 4, and 23;
October 20, 21, 23, 24;
November 4, 10, 11, 15, 16, 29, and 30; and
December 7, 8, 11, 12, 14, 15, 22, and 23.

Despite the apparent simplicity of this task, it is difficult to choose the time intervals suitable for weak-flare observations. Temporal overlapping of events neighboring strong flares impedes the selection of a suffiently large number of events for the analysis. Of course, it would be nice to analyze the maximum possible numer of events. However, the solar activity in 1996–1997 grew to a level which made data-aqsuisition challenging. In this respect, 1995 was the most favourable period to observe low-flux bursts. We can say confidently that carefully selected data constitute a representative sample to correctly represent the results.

In total, 296 bursts were observed during the above-mentioned time interval, including:
16 class-B flares with fluxes $10^{-7}$–$10^{-6}$ $W/m^2$;
139 class-A flares $10^{-8}$–$10^{-7}$ $W/m^2$; and
141 class-0 flares $10^{-9}$–$10^{-8}$ $W/m^2$.

Daily variations in the thermal background level in each instrument channel were small for all observations. The rms error was $\sigma \leq 0.3$ (Fig. 4a).

The arithmetic mean of the thermal background values at the beginning and end of a day was accepted as the mean thermal background $\alpha_1$ for that day. The thermal background is an X-ray flux that is registered by solar X-ray photometer, without solar events. The beginning and the end of a significant event (an X-ray flare) were identified based on variations in the daily mean of the X-ray flux value relatively to the thermal background value according to the following criterion: an event was triggered if the X-ray flux value had been increasing monotonously towards the thermal background daily-mean during 10–20 s up to $\alpha_2 \geq \alpha_1 + 0.5$. The event was traced till the time when, upon peaking, the daily-mean thermal background reached the initial daily average value $\alpha_1$. This was considered a sign for the end of the enhanced-count-rate event. In addition, according to our criteria, a significant event should have a well-pronounced peak whose amplitude exceeds $3\sigma$ in terms of the daily-mean thermal background. The durations of the onset and decay phases were determined based on the beginning, peak, and end times of a significant event.

Hereafter we discuss the GOES and INTERBALL data (Fig. 3). The GOES data were used to confirm the weak-burst detections. Good agreement between the data sets of two independent instruments gives us the hope, that solar events under analysis are reliable. The GOES data were taken from the Internet site [1].

In Fig 3, the count rate of the RF-15I-2 detector (the INTERBALL project) is plotted versus the power of the X-ray flares (the GOES project). The vertical dashed line corresponds to $10^{-8}$ W/m². It is clearly seen that the 0-class flares populate the distribution "tail" to the left of the dashed line below $10^{-8}$ W/m².

According to the work by [13] devoted to analyzing the energy distribution 15,000 solar flares observed in 1978–1979, the number of events grows sharply as the flare energy decreases down to $10^{28}$ erg, but begins to decrease with the further decrease in the flare energy. Hence, a maximum in the distribution noted in [13] cannot be attributed to selection effects. However, this statement is not solidly justified and the existence of the low-energy maximum in the event distribution remains an open question. In our opinion, the lower limit in the solar-flare energy distribution can exist, otherwise the energy distribution of low-flux X-ray solar flares would grow without limits with a decrease in the flare power. Note that this statement has also not been justified to date. We only see that the numbers of class-A and -0 events are fairly similar (139 and 141, respectively). All the low-flux events observed in the course of the INTERBALL project are confirmed by GOES observations.

Figure 4 shows the intensity time profiles of the solar X-ray microflares detected with the RF-15I-2 instrument in the energy range 2-3 keV from 0:00 to 12:30 UT on December 15, 1995. Class-0 flares are marked with arrows and figures.

The amplitudes of the events observed on December 15 range 1.5–2.5 count/s. This is above the $3\sigma$ level (see Fig.4a). The powers of class-0 flares observed on December 15 range 5.7 x $10^{-9}$ to 8.2 x $10^{-9}$ W/m². The parameters of an A class-A flare observed on December 15 are as follows: the beginning is at 0:17:37 UT, the peak is at 0:27:17 UT; the amplitude is 5.8 count/s against thermal background of 8 count/s, and the flare power is 1.7 x $10^{-8}$ W/m².

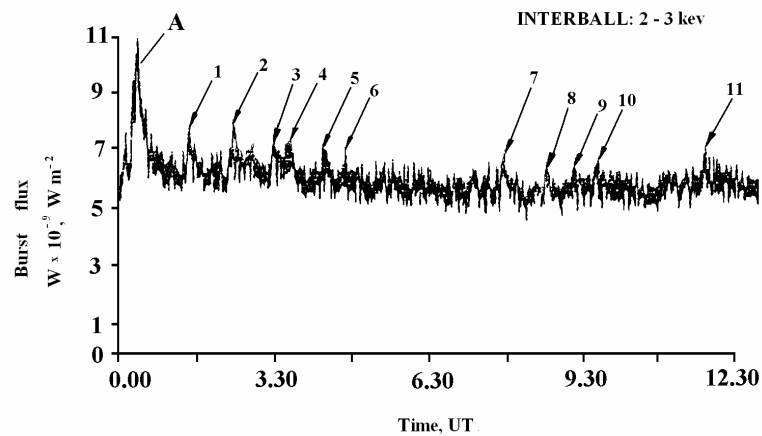

Fig. 4. Solar low-intensity X-ray flares detected in the 2-3 keV channel on December 15, 1995 since 0:00 UT.

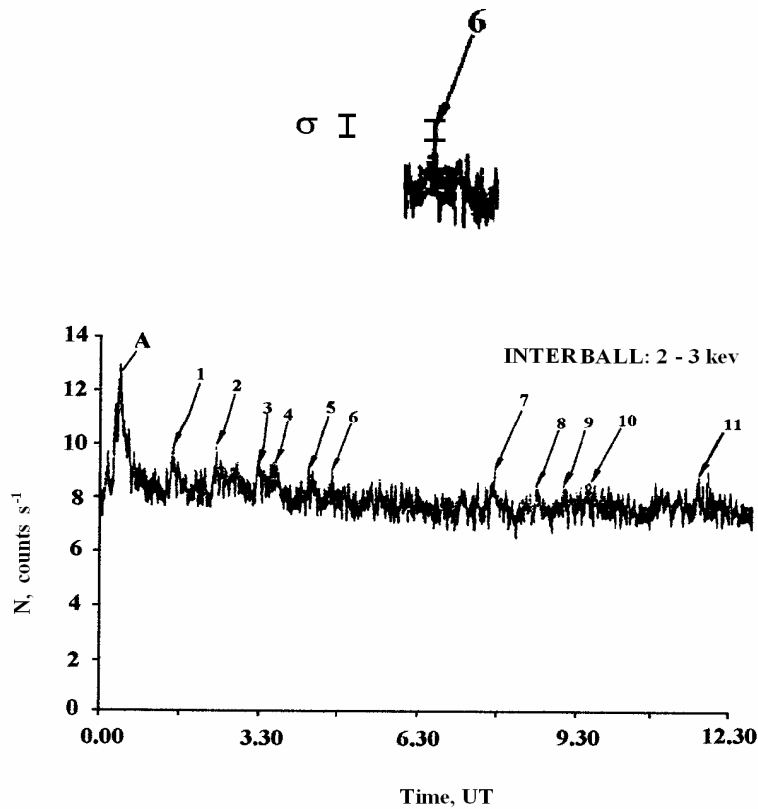

Fig.4a. Amplitudes of solar X-ray microflares in the energy range 2-15 keV were above the 3σ level.

Intensity time profiles of solar X-ray microflares detected by the GOES and INTERBALL projects on December 15, 1995 from 0:00 to 12:30 UT are compared in Fig. 5. The GOES time resolution is 1 min. Class-0 events are marked by the arrows and identified by the corresponding numbers. Right coincidence of temporary profiles is clearly seen.

The mean characteristics of low-intensity class-0 flares for the studied time interval are listed in Table 2. It is seen in this table that the mean flare duration is 5 □ 2 min, the mean peak flux exceeds the background by 2.2 count/s, and the mean flare power is $7 \times 10^{-9}$ W/m$^2$. In addition, solar events with the lowest duration and power meanings were registered on the 14th and the 22nd of December in 1995.

The burst of December 14 had the minimal duration of 30 s. The beginning and end times were 18:30:39 and 18:30:51 UT, respectively. The rise time was 12 s. The peak flux above background was 1.2 count/s. The background level was 8 count/s. The burst flux was $6.6 \times 10^{-9}$ W/m$^2$.

The burst of December 22 had the minimal flare intensity $4.5 \times 10^{-9}$ W/m$^2$. The beginning, end, and peak times were 18:43:24, 18:46:06 and 18:45:02 UT, respectively. The burst duration was $2^m 42^s$. The peak flux above background was 1.2 count/s. The background level was 6 count/s.

Conclusions:

1. 139 class-A and 141 class-0 flares have been studied from September 2 to December 23, 1995. A class 0 X-ray burst with the following parameters was observed in the energy range 2–15 keV:

- a duration of 30–300 s,
- a flux of $4.5 \times 10^{-9} – 10^{-8}$ W/m$^2$,
- a peak flux 1–5 count/s above background
- a thermal background level of 6–10 count/s.

  2. The total numbers of class-A and -O flares are similar.

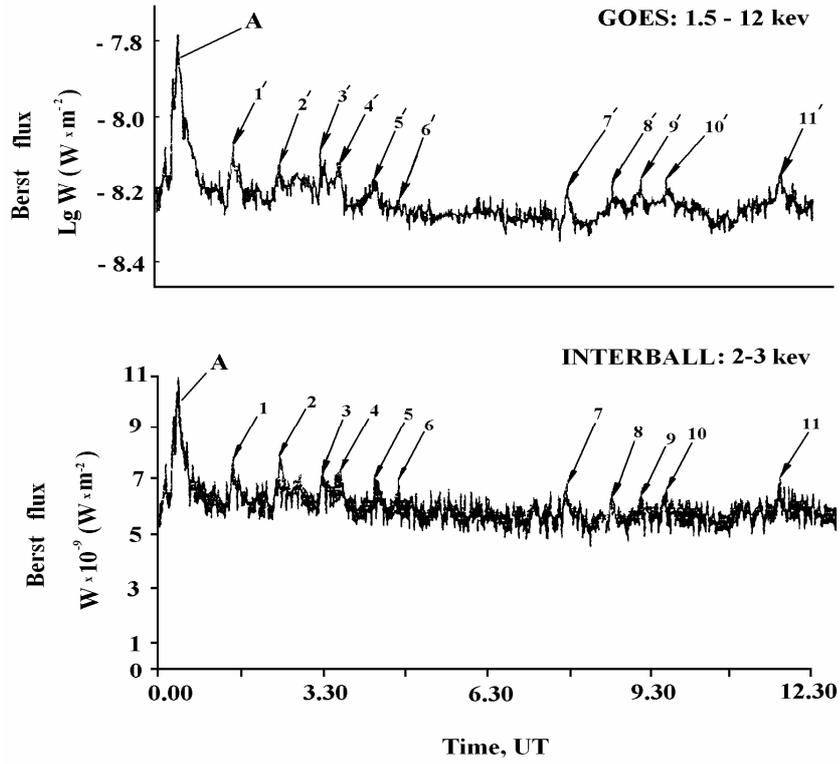

Fig.5. Comparison of the time profiles of solar X-ray microflares detected by the GOES and INTERBALL projects on December 15, 1995 from 0:00 to 12:30 UT.

Table 2. Mean parameters for weak class-0 solar soft X-ray bursts

| Date, 1995 | Mean duration | Mean rise time | Mean maximum intensity above background | Mean background level | Mean burst flux, W/m$^2$ |
|---|---|---|---|---|---|
| November 11 | $8^m44^s$ | $4^m48^s$ | 3.65 | 10 | 2.04 x 10$^{-9}$ |
| November 30 | $8^m20^s$ | $3^m42^s$ | 3.15 | 7 | 8.91 x 10$^{-9}$ |
| December 7 | $4^m07^s$ | $2^m04^s$ | 1.4 | 8 | 7.7 x 10$^{-9}$ |
| December 8 | $4^m04^s$ | $1^m34^s$ | 1.5 | 7 | 5.95 x 10$^{-9}$ |
| December 12 | $1^m47^s$ | $1^m24^s$ | 1.1 | 9 | 9.95 x 10$^{-9}$ |
| December 14 | $2^m23^s$ | $1^m20^s$ | 1.2 | 8 | 7.8 x 10$^{-9}$ |
| December 15 | $3^m42^s$ | $1^m12^s$ | 2.2 | 7.5 | 6.93 x 10$^{-9}$ |
| December 22 | $5^m07^s$ | $2^m42^s$ | 1.6 | 6.6 | 5.8 x 10$^{-9}$ |
| December 23 | $6^m06^s$ | $2^m26^s$ | 1.5 | 9 | 9.5 x 10$^{-9}$ |
| Generalized means | | | | | |
| | $4^m58^s$ | $2^m22^s$ | 2.2 | 8 | 7.1 x 10$^{-9}$ |

## 4. Microflares As A Phase Of A Solar Flare

Low-flux solar X-ray bursts discussed in this work have are characterized by total energy releases of $10^{25}$–$10^{26}$ erg. It should be noted that such events were observed in the very narrow photon energy range 2–5 keV. Within the framework of the classification proposed in [5], class-0 events can be considered microflares with an upper limit of the energy release ranging $10^{26}$ to $10^{27}$ erg. Figure 1 from [5] shows a gap in the data exactly in this range. Concerning events with energy releases $\leq 10^{25}$ erg, these so-called nanoflares were pointed out above to belong to a specific class of very-low-intensity solar events. They are observed only in the ultraviolet and actually are not flares. Consequently, only class-0 events with energy releases above $10^{25}$ erg can be considered microflares.

Revealing the features of the process underlying a microflare was based on the fact that it is possible to conclude on the thermal or nonthermal origin of a considered X-ray burst according to its time profile and photon energy range. The data collected and analyzed by S. Krucker in the RHESSI experiment [14] included 1000 soft X-ray microflares. Their time profiles were compared with the spectra. An example is shown in Fig.6 from [14].

According to the obtained data, the X-ray emission from a flare is of non-thermal origin with the probability of 90% if the time profile of this X-ray flare has a relatively short-term onset phase lasting no longer than 10% of the total flare duration and a longer-term decay phase. A fairly symmetric triangular time profile with both onset and decay phases lasting for about 50% of the flare duration or even a smoother and flatter time profile are indicative of the thermal origin of a flare. According to RHESSI data [14], 700 of 1000 studied time profiles were typical of the non-thermal origin of the X-ray flare emission. Our study is focused on very-low-intensity events observed, as noted above, in a very narrow photon energy range. Hence, it is hardly possible to construct energy spectra of such events in wide photon energy range. Meanwhile, it is possible to conclude on the mechanism responsible for the observed X-ray emission of the considered microflares using RHESSI data [14] on the time profiles of these events.

Figure 7 shows the solar X-ray bursts detected by the RF-15I-2 instrument in the 2–3 keV channel on December 8, 1995 from 01:30 to 06:00 UT. For comparison, the GOES data [1] obtained within the same time interval are also shown in this figure. The time resolution of the data by the RF-15I-2 detector in the 2–3 keV channel is 2 s. The GOES time resolution is 1 min.

In particular, a class-A flare with the beginning, peak, and end times 04:26:47, 04:28:49, and 04:45:07 UT, respectively, is seen in Fig. 7. The power of this event is $1.2 \times 10^{-8}$ W/m$^2$. Such features of its time profile as a relatively short-term leading edge with a duration of $2^m 2^s$, a pronounced peak, and a longer-term (in comparison with the leading edge) decay phase with a duration of $16^m 47^s$ are typical of X-ray bursts generated by the bremsstrahlung mechanism. The powers of all other events observed on December 8, 1995 from 01:30 to 06:00 UT were less than $10^{-8}$ W/m$^2$. These events marked with arrows and figures from 1 to 7 in Fig.7 belong to the class 0.

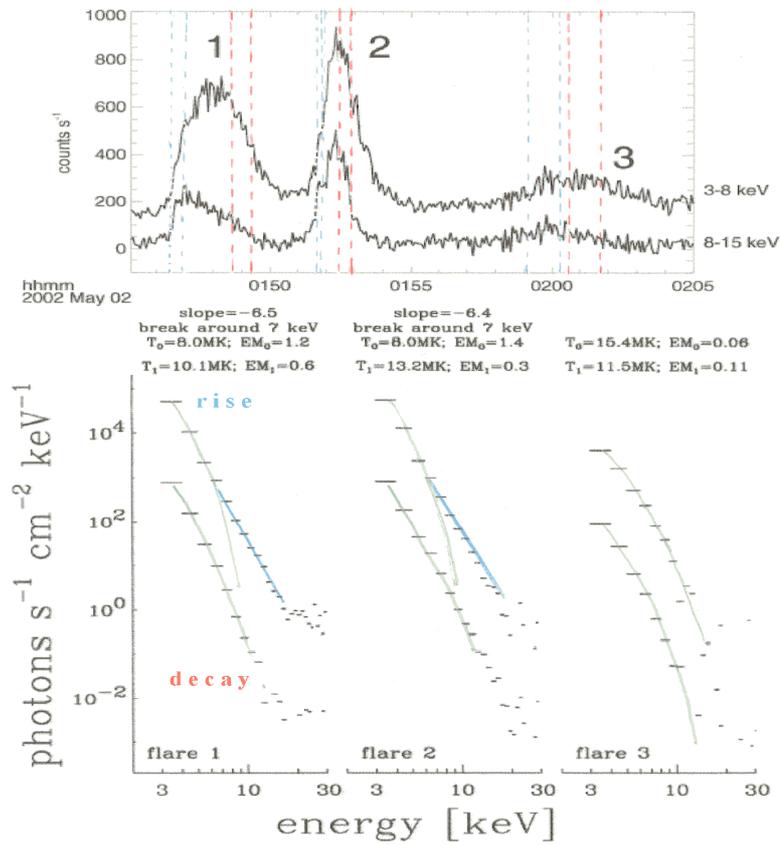

Fig.6. Microflare time profiles and spectra.

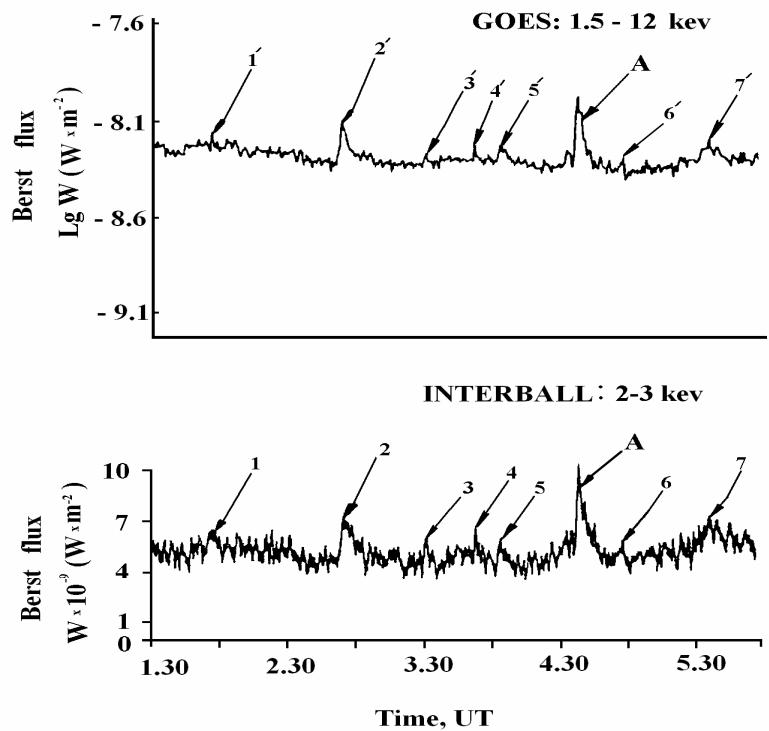

Fig. 7. Weak bursts in the solar X-ray emission registered in GOES and INTERBALL projects on the 8th of December in 1995 within the period from 01.30 UT up to 06.00 UT.

The time profile of interesting event 2 shows a short-term onset phase and relatively long-term decay phase, i.e., is typical of non-thermal X-ray emission. The beginning, peak, and end times of this event are 02:46:05, 02:46:51, and 02:57:57 UT, respectively. The onset and decay phases lasted $46^s$ and $11^m06^s$, repectively. The peak power of the flare is $6.8 \times 10^{-9}$ W/m$^2$. Such an event can be classified a class-0 microflare. Note that the time profiles of class-0 microflares are similar to the time profiles of higher-power events.

It is possible to assume that bremsstrahlung accompanying acceleration of electrons takes place on the "microlevel" already at the initial stage of energy release in a current sheet [9]. Note that, most likely, the current sheet has fragmented filamentary microstructure in which numerous irregularities can be formed, so that numerous microtears in individual current-sheet filaments may occur. Each microtear gives rise to a local electric field and electron acceleration. As a result, X-ray emission is generated.

In general, such a mechanism is identical to the processes described in [8], [15], which take place in higher-power flares. The contributions by the thermal and bremsstrahlung components to the burst can be different depending on the electron energy [16] and the structure of plasma-magnetic environment. Hence, both thermal and nonthermal processes coexist at the initial stage of flare formation. These processes are observed as class-0 microflares.

The scales of flaring events can be compared using the parameters of the highest- and lowest-power classes X and 0, respectively. It is seen in Table 3 that such parameters of class-X and -0 flares as the power, the thermal-background level, the photon energy range of the X-ray emission differ by one order of magnitude; the duration and the typical size of energy-release region, by 2–3 orders; and the excess of the peak flare intensity over the thermal background, by 2–4 orders of magnitude.

Table 3.

| Flare parameters | Class X | Class 0 |
|---|---|---|
| Power | $10^{-4}$–$10^{-3}$ W/m$^2$ | $10^{-9}$–$10^{-8}$ W/m$^2$ |
| Duration | 1.5–7 hours | 30–300 seconds |
| Thermal-background level | 9–60 count/s | 6–10 count/s |
| Excess of the peak flare intensity over the thermal background | 100–30000 count/s | 1–5 count/s |
| Photon energy range of the X-ray emission | 2–240 keV | 2–15 keV |
| Typical size of the energy-release region | $10^8$ cm | $10^5$–$10^6$ cm |

Similar to flare 2, the X-ray emission in flares 3, 5, and 6 (Fig. 7) can also be of bremsstrahlung origin. The beginning, end, and peak times of flare 3 are 03:29:36, 03:38:02, and 03:30:32UT, respectively, the onset- and decay-phase durations are $56^s$ and $7^m30^s$, respectively, and the peak flare power is $5.8 \times 10^{-9}$ W/m$^2$. The beginning, end and peak times of flare 5 are 04:01:09, 04:16:49, and 04:02:31UT, respectively, the onset- and decay-phase durations are $1^m22^s$ and $14^m18^s$, respectively, and the peak flare power is $5.3 \times 10^{-9}$ W/m$^2$. The beginning, end, and peak times of flare 6 are 04:57:55, 04:59:41, and 04:58:27 UT, respectively, the onset- and decay-phase durations are $32^s$ and $1^m14^s$, respectively, and the peak flare power is $5.1 \times 10^{-9}$ W/m$^2$.

Note that similar to event 2, the time profiles of events 3, 5, and 6 (Fig. 7) show relatively long-term decay phases typical of bremsstrahlung X-ray emission.

Thus, events 3, 5, and 6 are most probably class-0 microflares.

Flares 1 and 4 have triangular-shaped time profiles. The time profile of event 7 is smoother and longer-term. Flare 7 has the following parameters: the beginning at 05:28:35 UT, the end at 05:40:49 UT, the maximum at 05:32:47 UT, the onset- and decay-phase durations are $4^m12^s$ and $08^m02^s$, respectively, and the peak flare power is $6.8 \times 10^{-9}$ W/m$^2$. Flares 1, 4, and 7 may be identified as microflares whose X-ray emission is mainly of thermal origin.

Figure 8 shows another three weak X-ray flares which can be classified as class-0 microflares. These events registered on December 8, 1995 within the time interval period from 06:30 UT to 08:30 UT are marked with arrows and figures from 8 to 10 in Fig.8.

The beginning, end, and peak times of flare 8 are 07:08:59, 07:11:43, and 07:10:19 UT, respectively. The peak X-ray power of flare 8 is $6.9 \times 10^{-9}$ W/m$^2$. Flare 9 has the following parameters: the beginning, end, and peak times are 07:20:21, 07:40:51, and 07:28:31 UT, respectively, and the peak X-ray power is $8.7 \times 10^{-9}$ W/m$^2$. Flare 10 has the following parameters: the beginning, end, and peak times are 07:42:53, 07:50:35, 07:44:59 UT, respectively, and the peak X-ray power is $8.1 \times 10^{-9}$ W/m$^2$.

The total, onset-, and decay-phase durations for event 9 were $27^m 30^s$, $08^m 10^s$, and $19^m 20^s$, respectively. The time profile of this microflare has a relatively short-term onset phase, a prominent maximum, and a long-term decay phase. This is typical of non-thermal X-ray emission.

The total, onset-, and decay-phase durations for microflare 10 were $27^m 42^s$, $07^m 06^s$, and $20^m 36^s$, respectively. Similarly to event 9, the X-ray emission from microflare 10 is of non-thermal origin.

Microflare 8 is an interesting event with a total duration of $02^m 44^s$. The onset and decay phases have approximately equal durations. It should be noted that this event was not registered in the GOES project. This fact is surprising since the power of this flare is fairly large. The time profiles of GOES X-ray flares contain only thermal background during this period, whereas flare 8 was registered at the same time by the RF-15I-2 photometer of the "Interball– Tail Probe" project (see Fig.8). The time profile of the X-ray emission from microflare 8 is trianglular-shaped. The X-ray emission from this microflare is also most probably of thermal origin.

Thus, a class-0 microflare can be considered a specific phase of solar-event formation, within which both thermal and nonthermal processes coexist. Evidently, class-0 microflares is an intermediate class of events between thermal enhancements of the solar X-ray emission forming typical background and flares *per se*. Class-0 microflare represents a local, sufficiently fast energy release with the above-mentioned observed parameters, a local disturbance against thermal background of composite nature. Such microflares can overlap under certain conditions and their superposition can be observed as a more prominent and long-term flare event.

According to the work [13] devoted to an analysis of the energy distribution of 15,000 solar flares observed in 1978–1979, the number of flares increases as the energy release decreases down to $10^{28}$ erg. However, as the energy release decreases below this level, the number of events begins to decrease to a certain nonzero value. In other words, there may exist some lower limit in this distribution. This lower limit is probably formed due to class-0 microflares. For example, Fig. 1 from [5] clearly shows breaks of the solar-event energy distributions determined from their X-ray emission in the photon energy ranges 2–15 and 15–240 keV, i.e., the energy distribution of flares and microflares tends to a constant merely in the domain of class-0 microflares. Of course, such breaks can be caused by threshold selection effects accompanying detection and data processing. So in our case, the studied class-0 microflares were also identified in the GOES data.

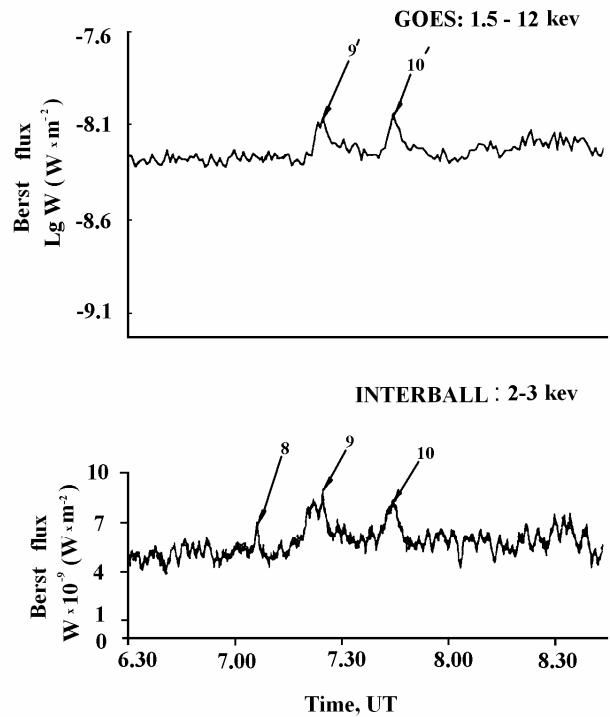

Fig.8. Weak solar X-ray bursts detected in GOES and INTERBALL projects on December 8, 1995 from 06:30 to 08:30 UT.

Good agreement of the data acquired by two independent instruments excludes possible instrumental and data-reduction effects and, to certain extent, confirms the validity of the considered solar events.

According to the above-referred work by [13], the relative amount of weak flares with lower energy releases is much bigger than the number of prominent flare events. Thus, microflares have the high chance to overlap and form more prominent flare events.

The existence of a maximum near $10^{28}$ erg in the solar-flare energy distribution can be explained by the more optimal physical conditions in the local active region where the energy can more easily be "stored" for a longer time in comparison with a larger active region with the higher energy-dissipation rate where a prominent flare event can arise.

Conclusions:

1. Class-0 X-ray flares registered in September–December, 1995 can be identified as microflares.

2. It is proposed that a flare is such an increase in the energy release in an active region that produces an ample amount of accelerated (heated) electrons which, in turn, generate certain minimum X-ray flux.

3. X-ray emission from microflares is generated by bremsstrahlung and/or thermal mechanisms.

4. Acceleration of electrons and the corresponding bremsstrahlung processes involved in the solar flare mechanism take place already during the initial phase of current-sheet formation, at some micro level which can correspond to the so-called elementary energy-release events observed as class-0 and -A X-ray flares.

## 5. Microflares And Thermal Background Of The Solar Corona
   Energy Distribution Of Solar Flares

The "Interball– Tail Probe" data were used to study the power distribution of solar X-ray bursts in the photon energy range 2–15 keV and the correlations of the daily mean peak fluxes of X-ray bursts of different classes with the daily mean thermal background of the solar corona.

These data imply that microflares are quasi-stationary events closely related to physical processes in the solar corona and capable of contributing significantly to the process of solar-corona heating (although not enough to explain the entire heating energy blance).

We analyzed microflares and the thermal X-ray background of the solar corona in the photon energy range 2–15 keV observed from September up to December, 1995. The observation period was chosen to avoid prominent solar flares and to unambiguously identify peak fluxes of lower-class X-ray events.

The microflare rate for the entire period of observation was 2 to 4 events per hour. The maximum rate corresponds to December 8, 1995 when 4 events per hour were observed on average during 21 hours of continuous observations.

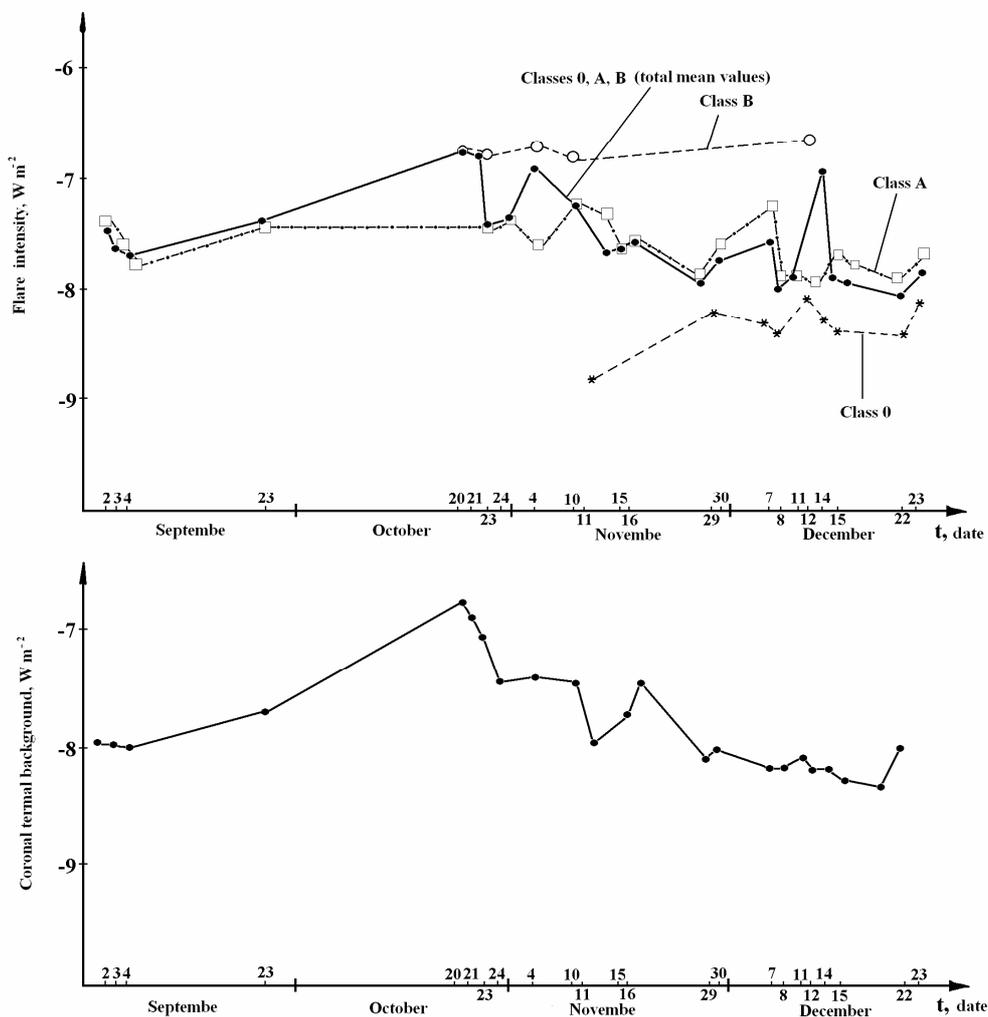

Fig.9 Comparison of the curve showing the total diurnal means of the peak X-ray fluxes of microflares of different classes and the curve showing the diurnal mean thermal X-ray background of the solar corona in the photon energy range 2–15 keV.

Table 4

| Average microflare power, W/m² | Number of microflares $N$ |
|---|---|
| Class B | |
| $3 \times 10^{-7}$ | 1 |
| $2 \times 10^{-7}$ | 5 |
| $1 \times 10^{-7}$ | 10 |
| Class A | |
| $9 \times 10^{-8}$ | 2 |
| $8 \times 10^{-8}$ | 3 |
| $7 \times 10^{-8}$ | 2 |
| $6 \times 10^{-8}$ | 7 |
| $5 \times 10^{-8}$ | 16 |
| $4 \times 10^{-8}$ | 12 |
| $3 \times 10^{-8}$ | 37 |
| $2 \times 10^{-8}$ | 39 |
| $1 \times 10^{-8}$ | 21 |
| Class 0 | |
| $9 \times 10^{-9}$ | 16 |
| $8 \times 10^{-9}$ | 15 |
| $7 \times 10^{-9}$ | 15 |
| $6 \times 10^{-9}$ | 61 |
| $5 \times 10^{-9}$ | 32 |
| $4 \times 10^{-9}$ | 1 |
| $2 \times 10^{-9}$ | 1 |

Comparative data on the detected microflares of different classes are presented in Fig. 9 showing the plots of the total daily peak fluxes of class-0, -A, and -B events and the plot of the daily mean thermal background of the solar corona during the observation period from September to December, 1995. Time is reckoned along the abscissa axis. The upper and lower panels in Fig. 9 show the logarithmic plots of the microflare X-ray flux and the daily mean thermal background in W/m², respectively. A certain number of microflares of different classes were observed in each day of the observation period. Based on the data for these events, the daily mean peak X-ray fluxes for each microflare class were evaluated. These values are used to plot the diagram in the figure. Hence, individual curves for class-0, -A, and -B microflares were obtained. In Fig. 9, class-0, -A, and -B curves are shown with the solid line connecting asterisks *, the dashdot line connecting squares □, and the dashed line connecting circles o, respectively. Moreover, the total diurnal means for all microflare classes (0, A, and B) were calculated for each observation day. This yielded the curve of the total diurnal mean X-ray peak flux shown with the black solid line in Fig. 9.

It is seen in Fig. 9 that the curve for class-A microflares and the curve for microflares of all classes are best correlated with the daily mean thermal background. The obtained data are indicative of closed relation between physical processes responsible for the microflares and the thermal solar-corona backround. Such a connection is realised most probably by virtue of plasma-magnetic structure of the solar corona and plasma-magnetic configurations of the active regions hosting microflares.

Table 4 lists the data on the number of microflares of different classes observed from September to December, 1995. The data are grouped according to the microflare subclasses. As a rule, the number of events decreases with increasing power, i.e., the number of higher-power events observed within certain interval of time is less than the number of lower-power events observed in the same period. It is seen in the table that such a trend typical of class-B events and partially for class-A events fails for class-0 events showing a peak of the rate for subclass 06.

This may be explained by the proximity of the power of a class-0 event to the thermal background and the absence of a distinct line between them [9].

It is seen in the right-hand part of Fig. 9 that the shape of the curve corresponding to class-0 microflares is fairly similar to that for the solar-corona thermal background.

In Fig. 10, the number $N$ of microflares is plotted versus the logarithm of the flare power $W$ in W/m$^2$. The black solid line is plotted with allowance for misroflare subclasses. The total numbers of class-A and -0 flares (139 and 141, respectively) are similar. This fact has already been noted in work by [6].

The dashed line in Fig. 10 is plotted without allowance for microflare distribution over subclasses. In this case, the vertical and horizontal axes, respectively, correspond to the total number of microflares of this class detected during the entire observation period and the peak X-ray power represented by the average X-ray flux of all microflares of this class within the observation period. The solid line peaks at about subclass 06 in the class-0 domain and drops to about constant minimum level at lower $W$.

To determine the total energy release of all class-0 flares, we took into account the average duration ($\approx$100 s) and average X-ray power ($7 \times 10^{-9}$ W/m$^2$) for solar X-ray events of this class. According to our estimates, the average entire energy release of all class-0 microfres is $E \approx 10^{25}$ erg. The total number of events in 1995 is estimated by extrapolating the results of 23 days of observations to the whole year. Since 159 class-C and 11 class-M large solar flares were observed in 1995, we evaluated the distribution of solar X-ray bursts for 1995.

Figure 11 shows the log–log energy-release distribution of X-ray solar bursts. Here, curve *1* is the distribution of X-ray bursts observed in 1995, i.e., during the solar-activity minimum, whereas curve *2* is the distribution of X-ray bursts observed in 1980, i.e., during a comparatively low maximum of the 20th solar-activity cycle according to the data from [17].

The break of the curve for 1980 is mainly explained by high background contributed by emission from solaractive regions. In our case, the background is extremely low, so the break of curve *1* is indicative of the actual deficiency in the number of low-flux events.

These results support our conclusion [6], [9] on the existence of the lower limit in the energy-release distribution of solar flares. This limit corresponds exactly to class-0 microflares closely related with the thermal background of the solar corona. Hence, the microflare energy release isan impotant energy source for solr-corona heating [18], although the entire solar-corona heating cannot be explained in this way since the number of microflares is small.

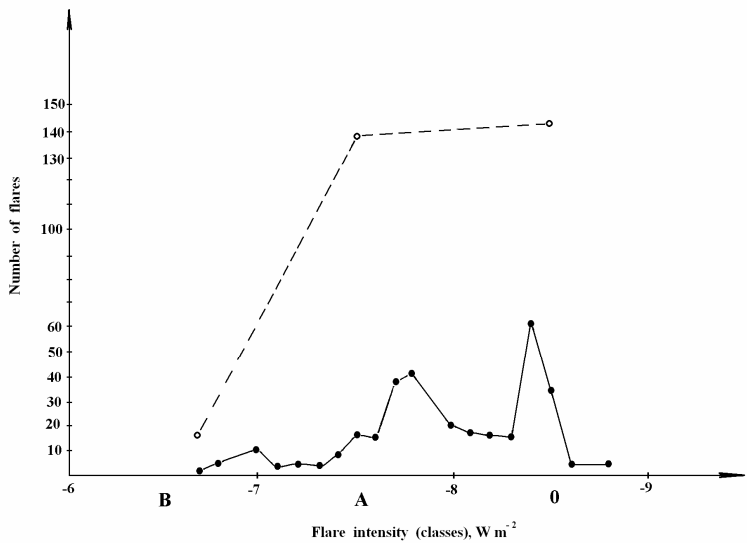

Fig. 10. Dependence of the number of microflares on their power for the chosen observation periods.

S.D.Christe [19], [3] comes to the same conclusion in his Ph.D. thesis "Hard X-ray Microflares" based on the RHESSI data [19] and in "RHESSI Microflare Statistics. I. Flare-Finding And Frequency Distributions" [3].

The number and rate distributions of RHESSI microflares in the GOES flux units are shown in Fig. 12 reproduced from [19].

Solar events with fluxes similar to typical background fluxes are shown with the solid line on the right plot. The peak of this distribution falls in the class-A domain, but the level of the solid line to the left of class A is fairly high. In other words, the events with the power of X-ray fluxes $<10^{-8}$ W/m$^2$ were observed. The left plot shows the rate of microflares. Here, the rate of events with fluxes similar to typical background fluxes is also shown with the solid line. It is seen in this plot that the rate of such events with GOES fluxes $<10^{-8}$ W/m$^2$ does not decrease down to zero, but remains at about constant level. This favors the conclusions on the existence of the lower limit in the energy-release distribution of microflares and the inability of the total energy release of such events to provide for the entire solar-corona heating.

Conclusions:

1. Analysis of the rates and total numbers of class-0 and -A flares and a comparison of these data with the rates of higher-power events reveal the existence of a lower limit in the solar-event distribution over the energy release. This limit is constituted by class-0 microflares.

2. Positive correlations of X-ray microflare time profiles and the thermal solar-corona background is discovered. This is indicative of a close relation between plasma-magnetic structure of the solar corona and plasma-magnetic configurations of the active areas hosting microflares.

3. Energy released via microflares is an essential contribution to the process of solar-corona heating. However, since the number of microflares is small, their total energy release is not sufficient to explain the entire solar-corona heating.

## 6. Summary

The main results are received:

1. Soft X-ray solar bursts are studied within the framework of the Interball–Tail Probe project. The low-intensity microflares observed in September–December, 1995 are analyzed. Weak bursts with powers up to $10^{-8}$ W/m$^2$ (class-0 events) in the X-ray range 2–15 keV were detected. Parameters of these microflares are determined.

2. It is proposed that a flare is such an increase in the energy release in an active region that produces an ample amount of accelerated (heated) electrons which, in turn, generate certain minimum X-ray flux.

3. X-ray emission from microflares is generated by bremsstrahlung and/or thermal mechanisms.

4. Acceleration of electrons and the corresponding bremsstrahlung processes involved in the solar flare mechanism take place already during the initial phase of current-sheet formation, at some micro level which can correspond to the so-called elementary energy-release events observed as class-0 and -A X-ray flares.

5. Analysis of the rates and total numbers of class-0 and -A flares and a comparison of these data with the rates of higher-power events reveal the existence of a lower limit in the solar-event distribution over the energy release.

6. Positive correlations of X-ray microflare time profiles and the thermal solar-corona background is discovered. This is indicative of a close relation between plasma-magnetic structure of the solar corona and plasma-magnetic configurations of the active areas hosting microflares.

7. Energy released via microflares is an essential contribution to the process of solar-corona heating. However, since the number of microflares is small, their total energy release is not sufficient to explain the entire solar-corona heating.

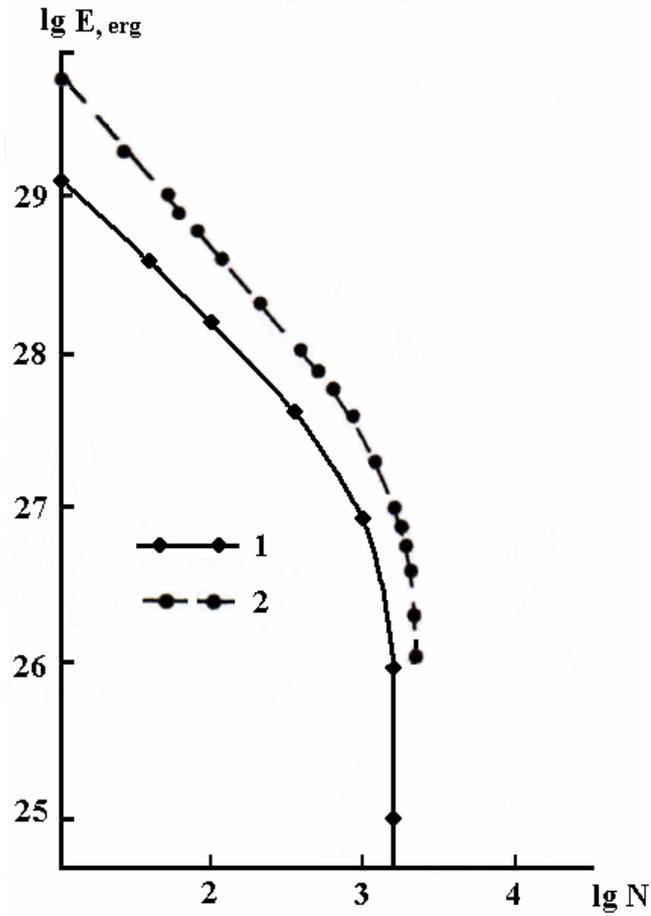

Fig 11. The energy distribution of X-ray solar flares in 1995 (1) and 1980 (2).

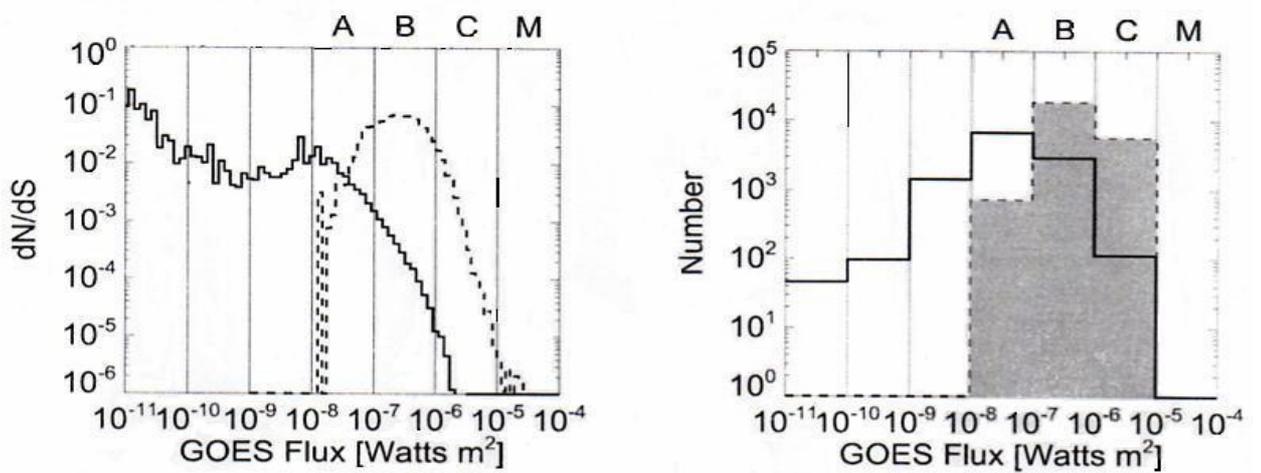

Fig.12. Number and rate distributions of RHESSI microflares in the GOES flux units.